\documentclass[sigconf]{acmart}
\usepackage{subfigure}

\def\BibTeX{{\rm B\kern-.05em{\sc i\kern-.025em b}\kern-.08emT\kern-.1667em\lower.7ex\hbox{E}\kern-.125emX}}
    
\usepackage{algorithm}
\usepackage{algorithmic}
\usepackage{booktabs} 
\usepackage{stfloats}
\usepackage{subfigure}
\usepackage{enumitem}
\usepackage[misc]{ifsym} 
\usepackage{bbding}
\usepackage{fancyhdr}

\renewcommand\footnotetextcopyrightpermission[1]{} 

\begin{document}
%
\title[Personalized Re-ranking for Improving Diversity]{Personalized Re-ranking for Improving Diversity in Live Recommender Systems}

%

\author{Yichao Wang, Xiangyu Zhang, Zhirong Liu, Zhenhua Dong, Xinhua Feng, Ruiming Tang, Xiuqiang He}
\affiliation{\institution{Noah's Ark Lab, Huawei, Shenzhen, China}}
\email{{wangyichao5, zhangxiangyu6, liuzhirong, dongzhenhua, fengxinhua1, tangruiming, hexiuqiang1}@huawei.com}

\renewcommand{\shorttitle}{Personalized Re-ranking for improving Diversity}
\renewcommand{\shortauthors}{Y. Wang, X. Zhang, Z. Liu, Z. Dong, X. Feng, R. Tang, X. He}
%

%
\begin{abstract}
Users of industrial recommender systems are normally suggested a list of items at one time. Ideally, such list-wise recommendation should provide diverse and relevant options to the users. However, in practice, list-wise recommendation is implemented as top-$N$ recommendation. Top-$N$ recommendation selects the first $N$ items from candidates to display. The list is generated by a ranking function, which is learned from labeled data to optimize accuracy. However, top-$N$ recommendation may lead to sub-optimal, as it focuses on accuracy of each individual item independently and overlooks mutual influence between items. Therefore, we propose a personalized re-ranking model for improving diversity of the recommendation list in real recommender systems. The proposed re-ranking model can be easily deployed as a follow-up component after any existing ranking function. The re-ranking model improves the diversity by employing \emph{personalized} Determinantal Point Process (DPP). DPP has been applied in some recommender systems to improve the diversity and increase the user engagement. However, DPP does not take into account the fact that users may have individual propensities to the diversity. To overcome such limitation, our re-ranking model proposes a \emph{personalized} DPP to model the trade-off between accuracy and diversity for each individual user. We implement and deploy the personalized DPP model on a large scale industrial recommender system. Experimental results on both offline and online demonstrate the efficiency of our proposed re-ranking model.

\end{abstract}

%
%

\ccsdesc[500]{Information systems~Recommender systems}

%
\keywords{Diversity, Re-ranking, Determinantal Point Processes, Recommender System}

%

%
\maketitle

\section{Introduction and Background}\label{sec:intro}

\begin{figure}[t]
 \centering
 \subfigure[a top-$N$ recommendation list without considering diversity]
 {\label{fig:similar_list}
 \includegraphics[trim=0 0 0 0,width =0.2\textwidth, height=0.32\textwidth]{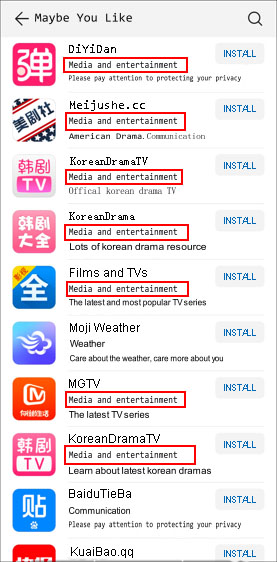}}
 \subfigure[a top-$N$ recommendation list, considering diversity]
 {\label{fig:diverse_list}
 \includegraphics[trim=0 0 0 0,width =0.2\textwidth, height=0.32\textwidth]{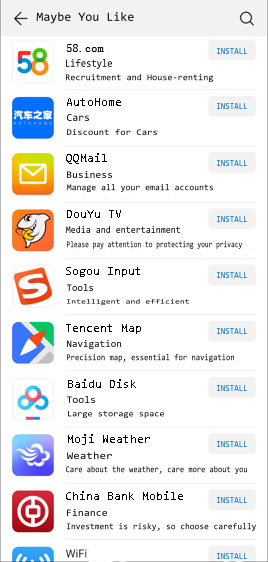}}
 \caption{Comparison of ranking list considering diversity or not.}
 \label{fig:list-example}
 \end{figure}

Recommender systems are powerful information filters for guiding users to find their interested items from gigantic and rapidly expanding pool of candidates, and they have taken more and more scenarios in our lives ~\cite{davidson2010youtube} ~\cite{schedl2015music} ~\cite{guo2018deepfm} ~\cite{li2008research}. Users in industrial recommender systems are normally recommended a list of items at one time. Ideally, such list-wise recommendation should provide diverse and relevant options to the users. Due to efficiency issue, many industrial recommender systems implement list-wise recommendation as Top-$N$ recommendation, which selects the first $N$ items from an ordered list. The ordered list is generated by a ranking function, which is learned from labeled data to optimize accuracy and produces a ranking score for each individual item. Such top-$N$ recommendation focuses on relevance of each individual item independently and overlooks mutual influence between items. As observed in~\cite{mcnee2006being}, recommending a list of items by such a method lead to sub-optimal performance of recommender systems, due to the following two aspects. On the one hand, ranking by relevance is likely to select multiple similar items in the list. However, it is highly possible that at most one of such similar items is needed by a user, while the others are redundant and waste the chance being displayed to the user. We take a real-world example from a mainstream App Store. As shown in Figure~\ref{fig:similar_list}, a top-$N$ recommendation list consists of multiple apps in the category of social community and video, because the ranking function learns that the user likes to chat with people and watch video. However, recommending multiple apps with the same functionality results wasting displayed chances and also degrading user experience. A more reasonable recommendation list should take diversity into account, as presented in Figure~\ref{fig:diverse_list}. On the other hand, focusing on relevance of items may lead to information isolation for the users~\cite{pariser2011filter}, which results in leaving fewer opportunities for exploring new items~\cite{nguyen2014exploring}. To address this problem, diversity~\cite{ziegler2005improving, zhang2008avoiding, bradley2001improving, adomavicius2012improving} has been imposed as a complement of accuracy, to model the mutual influence between items and therefore improve the effectiveness of list-wise recommendations. 

A multitude of approaches have been presented to generate the diverse recommendations. Industrial recommender systems are a very complicated framework so that we have to consider the ease of implementation and the risk of launching online when we deploy new models or components online. Therefore, considering diversity in recommendation, we aim to propose a component that is compatible with all the existing components, instead of replacing some of them. That is to say, we target ``diversity" as a re-ranking model, which can be easily deployed as a follow-up component after any existing ranking function. Some existing models exist, which treat ``diversity" as a re-ranking model, assuming a ranking of items is available. For example, Maximal Marginal Relevance (MMR) method in~\cite{carbonell1998use} selects one item at a time from a ranking list, which considers both of the relevance and the pair-wise similarity. Probabilistic models based on Determinantal Point Process (DPP) in~\cite{chen2017improving, kulesza2011k} consider the list-wise similarity among items through a kernel matrix, which consists of relevance and pair-wise similarity. Compared with the MMR-based model, DPP-based model can improve the diversity more efficiently without degrading the accuracy~\cite{chen2017improving}. However, we observe that an unrealistic assumption is made in DPP-based model: it is assumed that different users have the same propensity to the degree of diversity. We find some evidence from both literature and real-world data to call for different propensity to diversity with different individuals.

\begin{figure}
\centering
\includegraphics[height=1.5in, width=3in]{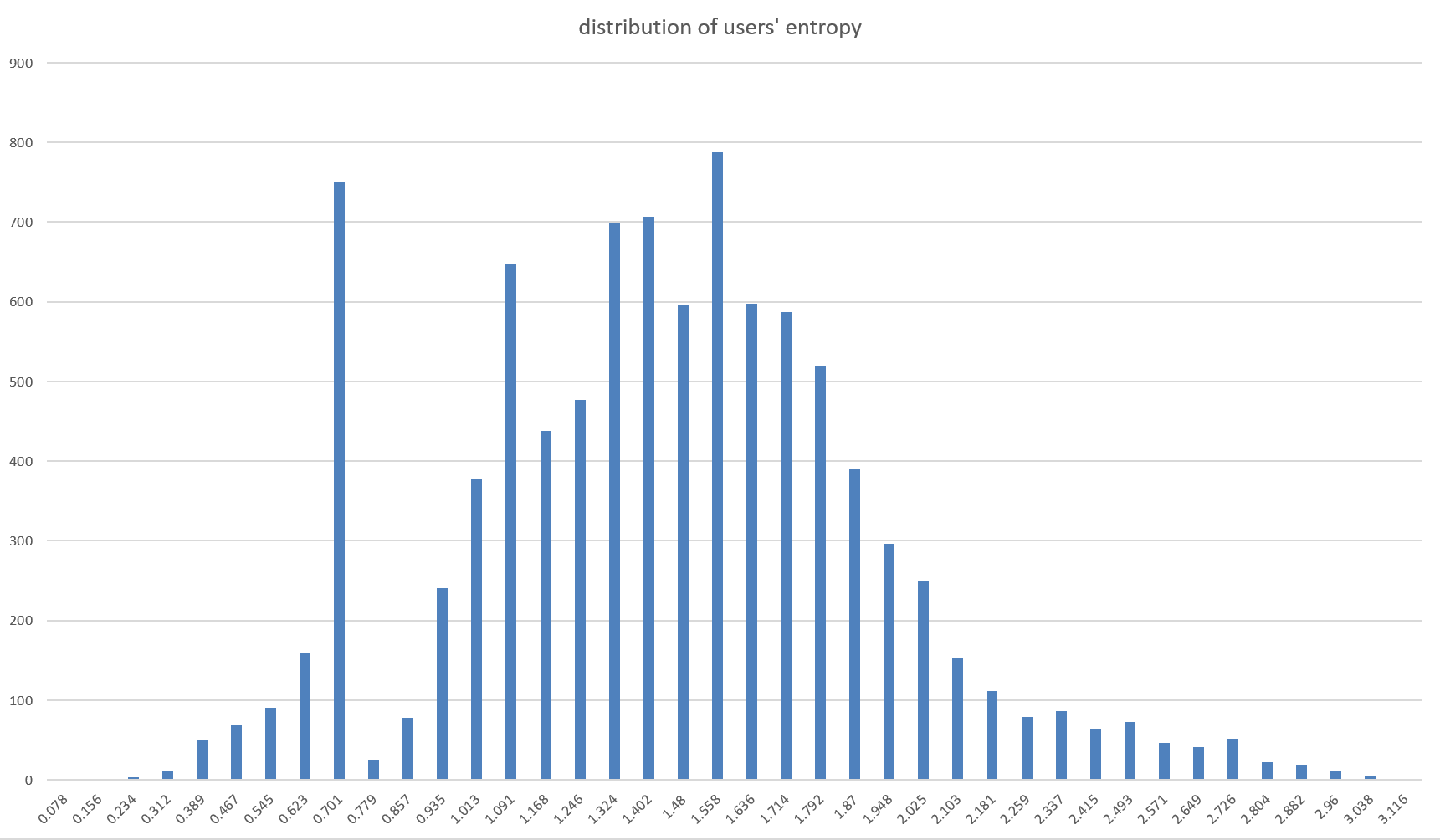}
\caption{The distribution of user's entropy over her download history collected from App Store with user consent}
\label{fig:entropy-distribution}
\end{figure}

Analyzing user behaviors from the same App Store with user consent, the result is presented in Figure~\ref{fig:entropy-distribution}. The figure presents the distribution of user's entropy over her download history. The $x$-axis represent the entropy value of the app category in her download history, while the $y$-axis represents the normalized population of this entropy value in the whole population of 10,000 users. It suggests that users' taste varies significantly. Users with large entropy values have a variety of interests over different categories of apps, while users with small entropy values focus on few categories of apps. It can be implied from this fact that different individuals have different propensity to diversity.

After seeking answer from industrial applications, we check the viewpoint of literature. \cite{chen2013personality} demonstrates that the personality traits of users significantly correlate with their behaviors in recommendation system. They take into account each users' personality trait based on a large scale user survey in \cite{wu2018personalizing, chen2016personality}, and present that different users have different propensity to the degree of diversity. Specifically, the users with narrow taste of items may expect more similar items in the recommendation list, while the users who have a variety of interests may expect more diverse items. Moreover, some researchers have proposed several methods to utilize the users' behaviors for building the personalized diversified recommendation. In \cite{di2014analysis}, the proposed algorithms focus on the users' propensity to diversity based on the different attributes of items, and re-rank the recommendation list by MMR methods. A pre-filtering approach proposed in \cite{eskandanian2017clustering} clusters the users into four groups according to each user's inclination to the diversity, and then apply the user based collaborative filtering algorithm for each group. These methods demonstrate the effectiveness of the personalized diversity with the offline experiment on some public datasets. 

However, \cite{di2014analysis,eskandanian2017clustering} personalize the diversity propensity on four user clusters, instead of on individuals, where each hyper-parameter in individual user cluster needs to be grid searched. As indicated from Figure~\ref{fig:entropy-distribution}, it is more reasonable to personalize the diversity propensity on individual users, as users' propensity to diversity varies significantly. However, it is impossible to extend~\cite{di2014analysis,eskandanian2017clustering} straightforwardly, personalizing diversity propensity from the granularity of user clusters to a much finer granularity of individual users, as searching for hyper-parameter for each user is impractical. Note that the number of users in an industry recommender system is normally tens or hundreds of millions. 

In this paper, we propose a personalized DPP model to improve the diversity of recommendation list, where the personalized granularity is of individual users. The hyper-parameter for each user is factorized to two factors: one is formulated by information entropy of a user's interaction history, while the other is commonly shared across all the users and tunable.

We summarize the main contributions of our study: 
\begin{itemize}
\item We propose a personalized re-ranking model for improving diversity of recommendation list, and it can be easily deployed as a follow-up component after ranking function.
\item The re-ranking model employs personalized DPP, where the penalization granularity is on individual users, instead of on user clusters in the literature. 
\item We conduct the experimental evaluations on an offline benchmark to show the superiority of our proposed re-ranking model. 
\item We deploy our proposed re-ranking model in a live recommender system and demonstrate the significant improvement for both of diversity and accuracy over baselines in online A/B test.
\end{itemize}

The rest of the paper is organized as follows: in Section~\ref{sec:methods}, we elaborate our re-ranking model in detail. We present our system architecture in live recommender systems in Section~\ref{sec:system}. Experimental setting and offline/online results are shown and discussed in Section~\ref{sec:experiment}. Finally, we give the conclusion in Section~\ref{sec:conclusion}.


\section{Re-ranking Model}
\label{sec:methods}
\subsection{DPP-based Re-ranking}

As studied in~\cite{chen2017improving}, DPP-based model is more effect and more efficient than other models such as  MMR-based model. Therefore, we choose to investigate how to apply DPP-based re-ranking model in our recommender system. In this section, we present DPP-based re-ranking model, and discuss its limitation, which motivates our personalized DPP-based re-ranking model in the next section. 

We summarize some key results about  from~\cite{chen2017improving, kulesza2011k,wilhelm2018practical}, for readers to better understand our model. A point process $\mathcal{P}$ on a set of items $\mathcal{M}=\{1, 2, ..., |\mathcal{M}|\}$ is a probability distribution on the powerset of $\mathcal{M}$. That is, $\forall Y\subseteq\mathcal{M}$, $\mathcal{P}$ assigns a probability $\mathcal{P}(Y)$, such that $\sum_{Y\subseteq \mathcal{M}}\mathcal{P}(Y) = 1$. It is stated in~\cite{wilhelm2018practical} that, finding the set $\max_{Y:|Y|=k,Y\subseteq \mathcal{M}}\mathcal{P}(Y)$ is a way of selecting a relevant and diverse subset of $k$ items from the whole item set $\mathcal{M}$. Furthermore, $\mathcal{P}$ can be compactly parameterized by a $\mathcal{M}\times \mathcal{M}$ positive semi-definite kernel matrix $L$, such that $\mathcal{P}(Y)\propto \det(L_{Y})$, where $\det(L)$ is the determinants of matrix $L$ and $L_{Y}$ is a submatrix of $L$ projected to only those rows and columns in $Y$. Therefore, find the set $\max_{Y:|Y|=k,Y\subseteq \mathcal{M}}\mathcal{P}(Y)$ is equivalent to finding the set $\max_{Y:|Y|=k,Y\subseteq \mathcal{M}}\det(L_{Y})$.

The positive semi-definite kernel matrix $L$ is defined as follows:
\begin{align}
\label{eq_kernel}
L_{ii} &= q_{i}^2 \\
\label{eq_kernel_2}
L_{ij} &= \alpha q_{i}q_{j}S_{ij}
\end{align}
where $q_{i} (i\in[1,|\mathcal{M}|])$ denotes relevance score of item $i$ generated from the ranking function, $S$ denotes a user-defined similarity matrix among the items, $\alpha$ is the hyper-parameter to trade-off relevance and diversity. 

As discussed before, we need to select a set of items $Y$ from the whole item set $\mathcal{M}$, such that
\begin{align}
\label{eq_optimize}
\max_{Y:|Y|=k,Y\subseteq \mathcal{M}}\det(L_{Y})
\end{align}

It is known as a NP-hard problem~\cite{wilhelm2018practical} with complexity $\mathcal{O}(\mathcal{C}_{|\mathcal{M}|}^{|Y|})$ to find the optimal set. To make DPP-based re-ranking model applicable in industrial recommender systems, we choose to use an efficient and effective approximation algorithm, Fast Greedy MAP Inference~\cite{chen2018fast}, to perform re-ranking in an acceptable latency. Such an approximation algorithm solves this combination optimization problem approximately in $\mathcal{O}(|Y|^{2}|\mathcal{M}|)$. Although theoretic lower bound is not provided in~\cite{chen2018fast}, online A/B test is conducted to demonstrate its superiority. 

To present it formally, we summarize DPP-based re-ranking model in Algorithm~\ref{alg:adpp}. Each round, FastGreedyMAP selects the one item greedily (as shown in Line 3 of Algorithm~\ref{alg:adpp}), which is to say, the selected item promotes the determinants of the updated submatrix most. Formally, it selects the item $y = argmax_{i\in \mathcal{M}}  \bigg ( \log \det(L_{Y\cup\{i\}}) - \log \det(L_{Y}) \bigg )$.

\begin{algorithm}
\caption{DPP-based Re-ranking model}
\label{alg:adpp}
\begin{algorithmic}[1]
\REQUIRE
    candidate items, $\mathcal{M}$; kernel matrix $L$; number of required items $k$;
\ENSURE 
    re-ranking list, Y; \\
\STATE \textbf{Initialize}: $Y=\Phi$;  
\WHILE{$|Y| < k$ and $|M| > 0$} 
\STATE $y$ = FastGreedyMAP($\mathcal{M}$, $L$, $k$)
\STATE $Y$ = $Y \cup y$ 
\STATE $\mathcal{M}$ = $\mathcal{M} - y$
\ENDWHILE
\RETURN $Y$
\end{algorithmic}
\end{algorithm}

\subsection{Personalized DPP}

In DPP, $\alpha$ is a tunable hyper-parameter to balance the trade-off between relevance and diversity. DPP assumes every individual have the same propensity to the degree of diversity, as the same $\alpha$ value is applied when constructing the kernel matrix $L$, which is shared when performing re-ranking for all users. However, as we discussed in Section~\ref{sec:intro}, different individuals have different propensity to diversity, so that personalization is needed in DPP.

A straightforward way to implement personalization in DPP is setting a unique hyper-parameter $\alpha_{u}$ for user $u$. Unfortunately, this approach is not practical, since the number of hyper-parameters $\alpha_{u}$'s is too large to be tuned individually. In this paper, we present an effect and efficient method to achieve personalized DPP (For short, we refer it as \emph{pDPP}). We factorize user-wise hyper-parameter $\alpha_{u}$ to two factors as 
\begin{align}
\label{eq_optimize1}
\alpha_{u} = f_{u}\times \alpha_{0}
\end{align}
where $\alpha_{0}$ is a tunable and shared hyper-parameter to trade-off relevance and diversity across all the users (which is of the same functionality as $\alpha$ in DPP) and $f_{u}$ is a user-wise factor representing diversity propensity of user $u$.

\begin{figure}
\centering
\includegraphics[height=2.2in, width=3in]{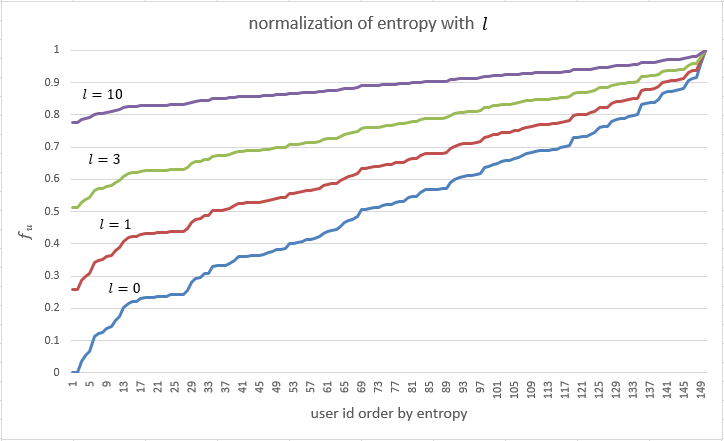}
\caption{Paraterized min-max normalized entropy with different hyper-parameter $l$ values.}
\label{fig:min-max-entropy}
\end{figure}

Next, we elaborate the intuition of defining $f_{u}$. As explained in a real-world example in Section~\ref{sec:intro}, users' diversity propensity can be reflected by their historical behavior. As one of the possible choices, Shannon entropy over the distribution of different genres\footnote{Our formulation can be extended easily by including other features of items.} of interacted items by the user is utilized, as 
\begin{align}
\label{eq_entropy}
\mathcal{H}(u)=-\sum_{g \in \mathcal{G}} P(g | u) \log (P(g | u))    
\end{align}
where $P(g|u)$ denotes the probability of user $u$ being interested in genre $g$, namely, one of user $u$'s interacted items being of genre $g$. As shown in~\cite{di2014analysis}, a user $u$ with higher $\mathcal{H}(u)$ has higher propensity of diversity and vice versa. Due to this intuition, we define $f_u$ as the normalized $\mathcal{H}(u)$. Formally, we propose to use a $parameterized$ $min$-$max$ $normalization$, as follows:

\begin{align}
\label{eq_fu_update}
    {f_{u} = \frac{\mathcal{H}(u) - \mathcal{H}_{\min} + l}{\mathcal{H}_{\max} - \mathcal{H}_{\min} + l}}\quad (l \geq 0)
\end{align}
where $\mathcal{H}_{\max} = \max_{u}\mathcal{H}(u)$ represent the maximal entropy value over all the users and $\mathcal{H}_{\min}$ denote the minimal value. The hyper-parameter $l$ controls the personalization degree of $f_{u}$ (and therefore $\alpha_{u}$). As shown in Figure~\ref{fig:min-max-entropy}, a larger $l$ value indicates less personalized $f_{u}$ values among all the users, e.g., when $l \xrightarrow{}\infty$, it can be seen that $f_{u} = 1$ and pDPP downgrades to DPP. In practice, we choose to use two special cases: when $l = 0$, $f_{u}$ is the standard $min$-$max$ normalized $\mathcal{H}(u)$; and when $l = \mathcal{H}_{\min}$, $f_{u}$ is the $max$ normalized $\mathcal{H}(u)$.

To summarize, pDPP is a personalized version of DPP without introducing extra hyper-parameters for tuning. Though the formulation is simple, the experiment results in Section~\ref{sec:experiment} demonstrate its effectiveness.

\section{System Implementation}\label{sec:system}

\subsection{Framework Modifications}

\begin{figure}
\centering
\includegraphics[height=2.2in, width=3in]{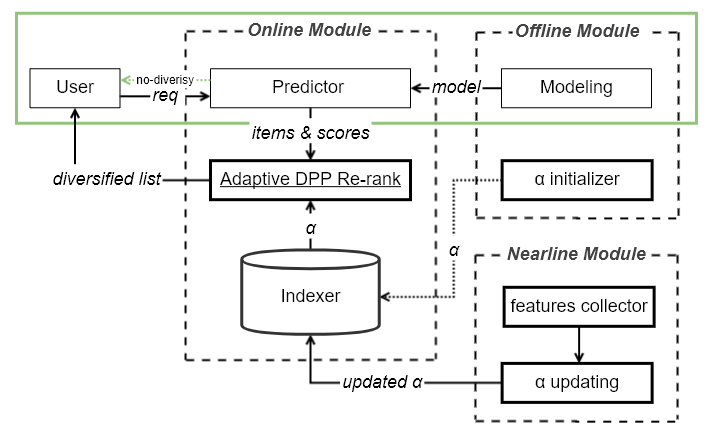}
\caption{The architecture overview of a recommender system with pDPP re-ranking model. The pDPP re-ranking model can be integrated with the other parts easily, and is compatible with any advanced ranking function.}
\label{fig:framework}
\end{figure}

An overview of a recommender system with pDPP re-ranking model is shown in Figure~\ref{fig:framework}. We first present the modules without considering the re-ranking model (which is surrounded in green box) and then illustrate how to adapt these modules with pDPP.

The architecture of a recommender system consists of three modules. (1) \emph{Offline training module} processes user-item interaction data, extracts features (user features, item features and context features), trains model and uploads the model. (2) \emph{Online prediction module} receives users' request and returns a list of items. There are usually two steps in this module, namely \emph{retrieval} and \emph{ranking}. Since there may be over millions of items, it is impossible to score every item within a required latency (often within tens of millisecond). The retrieval step returns a short list of items (often hundreds or thousands) of items that is suitable for the user under such context. After reducing the size of candidates, the ranking step computes relevance scores for individual items using the offline trained model. (3) \emph{Nearline updating module}, which updates user features, item features and even the offline trained models with real-time interaction data.

Our proposed pDPP re-ranking model can be integrated into the above architecture easily. Next, we will elaborate how to adapt the three modules in the framework, to deploy this re-ranking model.
\begin{itemize}
\item In offline training module, $\alpha$ $initializer$ computes $\alpha_{u}$ value for individual user $u$ and uploads such values to online $Indexer$.
\item In online prediction module, given the relevance scores of candidate items computed by any ranking function and the personalized $\alpha_{u}$ value from online $Indexer$, pDPP re-ranking model generates the final recommendation list, considering both relevance and diversity.
\item In nearline updating module, personalized $\alpha_{u}$ values are updated based on the real-time user-item interaction data, and the updated $\alpha_{u}$ values are sent to online $Indexer$.
\end{itemize}

Developing accurate ranking function is an essential research topic and attracts many researchers from both academia and industry. As can be seen, our pDPP re-ranking model is compatible with any advanced ranking function, without any modification on such ranking function.



\subsection{Practical Issues}

To help readers better understand and implement our model in their recommender systems, we summarize several practical issues which should be noticed in real-world applications. 

\begin{itemize}
    \item In research work such as~\cite{chen2018fast}, the kernel matrix $L$ is pre-computed and stored in memory, as shown in Algorithm~\ref{alg:adpp}. However, such a method cannot be performed in a real-world recommender system, due to the following two reasons. Firstly, the relevance score $q_i$'s, computed by a ranking function, are personalized and real-time updated. Such industrial-style ranking function makes different relevance scores of individual users to the same item, and furthermore, the relevance score of a user-item pair may be updated in a few seconds as the user feature may be changed. Secondly, our pDPP model has a personalized factor $f_{u}$ when constructing $L$ so that different users have different $L$. We need a huge amount of time and storage resources to handle such $L$'s if we need to pre-compute and store them. Due to these two reasons, we compute the personalized kernel matrix $L$ for a user on-the-fly when this user trigger the request to our recommender system.
    
    \item In our experiments, we tried two different approaches to construct the similarity matrix $S$: one utilizes item features and the other uses user-item interaction information. The method with user-item interaction performs slightly worse than with the other. The reason may due to the fact that user-item interactions are usually very sparse which makes the item representations based on such information not very reliable. No matter which approach are used, we find that the performance is better when we normalize $S_{ij}$ in $[0,1]$.
    \item Cold start problem is one of the common challenges in recommender systems. In our system, we set $\alpha_{u} = \alpha_{0}$ if $u$ is a new user. Moreover, users with only few interactions are also regarded as new users by our system. We make such a decision because $\alpha_{0}$ is a relatively safe value for exploration, while balancing the trade-off between relevance and diversity.
\end{itemize}

\section{EXPERIMENTAL EVALUATION}
\label{sec:experiment}
To demonstrate the superiority of our pDPP-based re-ranking model, we firstly design offline experiment on two datasets to compare the relevance and diversity of recommendation result of our model with that of baselines. Furthermore, we deploy our model on a live recommender system, to validate its effectiveness in an industry application. In this section, we will present the experiment details and analyze the results in terms of offline and online evaluation, respectively.

\subsection{Offline Evaluation}

\subsubsection{Datasets}

For offline evaluation, we prepare two datasets. Besides MovieLens, which is a benchmark in recommendation research community, we also collect user-item interaction log from our commercial App Store. To help reproduce our experiment result, we firstly describe how we process such two datasets.



\emph{MovieLens 1M Dataset}
\footnote{\url{http://grouplens.org/datasets/movielens/1m/}} contains 1,000,209 anonymous ratings with approximately 3,900 movies rated by 6,040 users. As a traditional pre-processing by research work as~\cite{chen2017improving}, we eliminate the movies rated by less than 10 users and the users rating less than 20 movies. We randomly split the ratings to two parts, where 70\% of the ratings are used for training, and 30\% are for testing. A samples with rating greater than or equal to 4 is treated as positive, otherwise negative. We perform item-based collaborating filtering as ranking function to predict the relevance score for each item. The similarity matrix $S$ is built based on the genres of movies, which is to say, $S_{ij} = 1$ if movie $i$ and $j$ are of the same genre. 

\emph{Company Dataset} is collected from our commercial App Store with user consent. This dataset contains approximately 100,000 download records from about 80,000 users in 8 consecutive days. The size of the whole item set is 7,000. Samples in the first 7 days are used for training, while samples in the last day are for testing. We consider all download records as the positive samples, and the others (i.e., the apps that are in the item set but not downloaded by a user) are negative ones. The similarity matrix $S$ is generated by the category of apps, i.e., $S_{ij} = 1$ if app $i$ and $j$ are of the same category.

\subsubsection{Baselines}

Two baselines are compared. The first baseline considers the ranking function for relevance and disregards re-ranking model for diversity, which is referred as \texttt{BASE}. The other baseline is the standard DPP for re-ranking, which is presented as \texttt{DPP}. Although MMR is a popular state-of-the-art method, we omit it here due to its inferiority compared to DPP~\cite{chen2017improving}. Our personalized DPP model for re-ranking is denoted as \texttt{pDPP}. Note that the ranking function\footnote{In MovieLens 1M Dataset, item-based CF is served as the ranking function; while in Company Dataset, a popular deep learning model is served.} utilized in \texttt{BASE}, \texttt{DPP} and \texttt{pDPP} keeps consistent, for fair comparision. The hyper-parameter $\alpha$ value in \texttt{DPP} and $\alpha_{0}$ value in \texttt{pDPP} are found by grid search. 


\subsubsection{Evaluation Metrics}

To compare the models comprehensively, we evaluate them from both \emph{relevance} and \emph{diversity} aspects of their recommendation results. \emph{Precision} is utilized to measure the relevance, which is defined as 
\begin{align}
\label{precision}
  precision &= \frac{\sum\nolimits_{u} |{R}_{u} \cap {T}_{u}|}{\sum\nolimits_{u} |{R}_{u}|}
\end{align}
where ${R}_{u}$ denotes recommendation list of user $u$, ${T}_{u}$ denotes download apps of user $u$ in test set. 

To measure the diversity, we adopt \emph{intra-list distance} ($ILD$) \cite{zhang2008avoiding}, which is defined as
\begin{align}
\label{ILD}
  ILD &= \underset{u}{avg} \underset{i,j \in {R}_{u}, i \neq j}{avg} (1-{S}_{ij})
\end{align}
$P@k$ and $ILD@k$ is the $precision$ and $ILD$ of the first $k$ item in the recommendation list. 

Moreover, as a third metric, we measure the balance of $precision$ and $ILD$ denoted as $avg(P@K,ILD@k)$ in \cite{di2014analysis, panniello2014comparing}, wherein the metrics are standardized to make the scales homogeneous.


\subsubsection{Experimental Results}
\begin{table}[h]
\caption {Experiment Results on MovieLens 1M Dataset} 
\label{table:movielens-result}
\begin{tabular}{cccc}
\toprule
Model  & $P@5$     & $ILD@5$    &    $avg(P@5,ILD@5)$  \\ 
\midrule
\texttt{BASE}           & 0.0420 & 0.5587 & 0.4000         \\ 
\texttt{DPP}$_{(\alpha=0.01)}$         & 0.0420 & 0.5623 & 0.4314 \\
\texttt{DPP}$_{(\alpha=0.02)}$           & 0.0419 & 0.5735 & 0.5035 \\ 
\texttt{DPP}$_{(\alpha=0.03)}$           & 0.0416 & 0.5928 & 0.3717 \\ 
\texttt{DPP}$_{(\alpha=0.04)}$           & 0.0415 & \textbf{0.6162}  & 0.5000         \\ 
\texttt{pDPP}$_{(l = \mathcal{H}_{\min})}$       & 0.0420 & 0.6015 & 0.7717 \\ 
\texttt{pDPP}$_{(l = 0)}$      & \textbf{0.0421} & 0.5938 & \textbf{0.8046} \\ 
\bottomrule
\end{tabular}
\end{table}
\vspace{-4mm}
Both offline experiments are performed for multiple times to ensure the results are statistically accurate. During each experiment, we randomly shuffle the data for training and test on the MovieLens Dataset and conduct consecutive experiments within different dates for Company Dataset.
Experiment results on MovieLens 1M Dataset with $k=5$ are shown in Table ~\ref{table:movielens-result}. Due to space limit, we omit the results with other $k$ values, but they are analogous. 

\texttt{DPP} model aims to balance the trade-off between accuracy and diversity, for which we can focus more on diversity (i.e., $ILD$ metric) by enlarging $\alpha$, but on the other hand, the relevance performance (i.e., $precision$ metric) will be degraded.
Compared with \texttt{BASE}, \texttt{DPP}$_{(\alpha=0.01)}$ achieves the same accuracy but better diversity. We select a reasonable range for $\alpha$, to avoid degrading the accuracy significantly, i.e., $\alpha = \{0.01, 0.02, 0.03, ..., 0.1\}$. Among such values, $\alpha = 0.04$ makes \texttt{DPP} performs the best in terms of $ILD@5$ while $\alpha = 0.02$ enables \texttt{DPP} achieves the best $avg(P@5,ILD@5)$. 

As expected, \texttt{pDPP} outperforms all the baselines in terms of $P@5$ and $avg(P@5,ILD@5)$, which demonstrates the superiority of modelling different propensity of diversity for individual users. Specially, \texttt{pDPP}$_{(l = 0)}$ performs best in terms of $P@5$ and $avg(P@5,ILD@5)$ and slightly decreases the performance of $ILD@5$ compared to \texttt{pDPP}$_{(l = \mathcal{H}_{\min})}$.
\begin{table}[h]
\caption {Experiment Results on Company Dataset} 
\label{table:company_result}
\begin{tabular}{cccc}
\toprule
Model              & $P@5$                  & $ILD@5$    & $avg(P@5,ILD@5)$ \\ 
\midrule
\texttt{BASE}                    & \textbf{0.0782} & 0.6554 & 0.5000            \\ 
\texttt{DPP}$_{(\alpha=0.1)}$                    & \textbf{0.0782}          & 0.6559 & 0.5014    \\ 
\texttt{DPP}$_{(\alpha=0.2)}$                    & \textbf{0.0782}          & 0.6599 & 0.5108     \\ 
\texttt{DPP}$_{(\alpha=0.3)}$                    & 0.0769          & 0.6695 & 0.4835    \\ 
\texttt{DPP}$_{(\alpha=0.4)}$                    & 0.0756          & 0.6775   & 0.4525    \\ 
\texttt{DPP}$_{(\alpha=0.5)}$                    & 0.0769          & 0.6865 & 0.5239    \\ 
\texttt{DPP}$_{(\alpha=0.6)}$                    & 0.0769          & 0.7048 & 0.5672    \\ 
\texttt{DPP}$_{(\alpha=0.7)}$                    & 0.0744          & 0.7160 & 0.4937    \\ 
\texttt{DPP}$_{(\alpha=0.8)}$                    & 0.0744          & 0.7301 & 0.5272    \\ 
\texttt{DPP}$_{(\alpha=0.9)}$                    & 0.0705          & 0.7682 & 0.4676    \\ 
\texttt{DPP}$_{(\alpha=1.0)}$                      & 0.0654          & \textbf{0.8662} & 0.5000       \\ 
\texttt{pDPP}$_{(l = \mathcal{H}_{\min})}$       & \textbf{0.0782}               & 0.7025   & 0.6118    \\ 
\texttt{pDPP}$_{(l = 0)}$ & \textbf{0.0782}             & 0.7051 & \textbf{0.6179}    \\ 
\bottomrule
\end{tabular}
\end{table}

The results on Company Dataset are shown in Table ~\ref{table:company_result}. Similar to the experiments on MovieLens 1M Dataset, we select an appropriate range as $\alpha = \{0.1, 0.2, 0.3, ..., 1.0\}$. Adding DPP-based re-ranking model based on \texttt{BASE}, the \texttt{DPP} models improve the diversity while sacrificing the performance of relevance. Compared with the \texttt{BASE} and \texttt{DPP} models, \texttt{pDPP} models(with 0.6 as the $\alpha_{0}$) gain the best performance on $P@5$ and $avg(P@5,ILD@5)$. Remarkably, retaining exactly the same accuracy as \texttt{BASE},  \texttt{pDPP} methods improve the diversity significantly. Specifically, between \texttt{pDPP} family, the model with $l=0$ achieves better diversity than the one with $l = \mathcal{H}_{\min}$ while keeps the same accuracy performance.

\subsection{Online Evaluation}

As shown its superior balancing the trade-off between accuracy and diversity in offline evaluation, we deploy \texttt{pDPP} in a live recommender system to verify its effectiveness in an industry application. 

\subsubsection{Experiment setting}

For online evaluation, we conduct online A/B test. We compare three different families of models: \texttt{BASE}, \texttt{DPP} and \texttt{pDPP}. We randomly split all the users into hundreds of bins, each of which consists of more than 100,000 users. A bin of users are served by each of the three compared models.
In our live recommender system, the hyper-parameter $\alpha$ of DPP and $\alpha_{0}$ of pDPP are set to $0.6$ as the performance of $avg(P@5,ILD@5)$ is the best when $\alpha=0.6$ in offline evaluation (as presented in Table~\ref{table:company_result}).

\subsubsection{Evaluation Metrics}

To compare the performance of these methods, we evaluate them on the basis two metrics of accuracy and one metric of diversity. The first accuracy metric that we measure is \emph{download ratio} ($DR$), defined as

\begin{align}
\label{down_rate}
  DR &= \frac{total\ number\ of\ downloads}{total\ number\ of\ impressions}.
\end{align}

Beyond that, we also measure the engagement of users. More specifically, we study \emph{average number of downloads} ($AD$) \emph{per user}, as

\begin{align}
\label{avg_user_downs}
  AD &= \frac{total\ number\ of\ downloads}{total\ number\ of\ users}
\end{align}

Besides these two accuracy metrics, we adopt $ILD$ to evaluate the diversity, the same as in offline evaluation.

\begin{table}[h]
\centering
\caption{Online A/B testing results}
\label{tabel:online_result}
\begin{tabular}{cccc} 
\toprule
 Model        & $DR$      & $AD$     & $ILD$      \\ 
\hline
\texttt{BASE} & -       & -       & -        \\
\texttt{DPP}  & +6.13\% & +2.83\% & +4.06\%  \\
\texttt{pDPP} & \textbf{+6.52\%} & \textbf{+3.54\%} & \textbf{+4.32\%}  \\
\bottomrule
\end{tabular}
\end{table}

\subsubsection{Online Performance} 

The results of A/B online test are shown in Table~\ref{tabel:online_result}. Considering the commercial concerns, we only present the relative improvement of \texttt{DPP} and \texttt{pDPP} over \texttt{BASE} model in terms of $DR$, $AD$ and $ILD$.  

We can observe that both \texttt{DPP} and \texttt{pDPP} perform significantly better than \texttt{BASE} in terms of all the three evaluation metrics. It suggests that improving diversity is able to boost the recommendation performance. Between \texttt{DPP} and \texttt{pDPP}, we observe that \texttt{pDPP} is superior than \texttt{DPP}, which indicates that personalized propensity to diversity is more suitable than identical propensity setting. 
We observe that the improvement of the \texttt{pDPP} over \texttt{DPP} is not as significant as that in offline evaluation. Through detailed analysis, we find that about 35\% of the users have only one download record in their behavior history so that it is hard to define their propensity to diversity under such circumstance, which may be one reason for the not-so-significant improvement. However, the daily turnover of our App Store is millions of dollars, therefore even such not-so-significant lift in $DR$ and $AD$ brings extra millions of dollars each year.


\section{CONCLUSIONS}
\label{sec:conclusion}

Recommender system which only focuses on \emph{accuracy} may lead to sub-optimal, as it too much emphasizes the \emph{accuracy} of each individual items and leads to presenting similar items. \emph{Diversity}, which has been studied to present the users with more diversified items, can be viewed as mutual influence among items. Therefore, combining accuracy and diversity in recommender system is a reasonable and convincing way to improve the performance. Furthermore, different users have different propensity to diversity, which requires personalized diversity. In this paper, we propose a personalized re-ranking model for improving the diversity of the recommendation list based on personalized DPP. This re-ranking model can be easily deployed as a follow-up component after any existing ranking function. The offline experiments over two real-world datasets and the online comparison through A/B testing in an industrial recommender system demonstrate the effectiveness of our proposed re-ranking model. 


%

\newpage
\bibliographystyle{unsrt}
\bibliography{diversity}

%

\end{document}